# Market-Based Variance of Market Portfolio and of Entire Market

Victor Olkhov

Independent, Moscow, Russia

victor.olkhov@gmail.com

ORCID: 0000-0003-0944-5113

## Abstract

We present the unified market-based description of returns and variances of the trades with shares of a particular security, of the trades with shares of all securities in the market, and of the trades with the market portfolio. We consider the investor who doesn't trade the shares of his portfolio he collected at time $t_0$ in the past. The investor observes the time series of the current trades with all securities made in the market during the averaging interval. The investor may convert these time series into the time series that model the trades with all securities as the trades with a single security and into the time series that model the trades with the market portfolio as the trades with a single security. That establishes the same description of the returns and variances of the trades with a single security, the trades with all securities in the market, and the market portfolio. We show that the market-based variance, which accounts for the impact of random change of the volumes of consecutive trades with securities, takes the form of Markowitz's (1952) portfolio variance if the volumes of consecutive trades with all market securities are assumed constant. That highlights that Markowitz's (1952) variance ignores the effects of random volumes of consecutive trades. We compare the market-based variances of the market portfolio and of the trades with all market securities, consider the importance of the duration of the averaging interval, and explain the economic obstacles that limit the accuracy of the predictions of the returns and variances at best by Gaussian distributions. The same methods describe the returns and variances of any portfolio and the trades with its securities.

This research received no support, specific grant, or financial assistance from funding agencies in the public, commercial, or nonprofit sectors. We welcome offers of substantial support.

# 1. Introduction

The importance of the market portfolio is generally accepted, and probably almost all problems related to the description and the use of market portfolios are well-studied (Athanasoulis and Shiller, 1997; Koutmos, 1997; Doeswijk, Lam, and Swinkels, 2014; Hollstein and Prokopczuk, 2023). We believe that the investors, portfolio managers, and researchers in finance are well acquainted with all features of market portfolios. However, we draw attention to this well-known topic and study a particular problem that determines the returns and the variances of the market portfolios and their interference with the returns and the variances of the trades with all securities in the market.

In his classical paper Markowitz (1952) derived the expression of the portfolio variance $\Theta_M(t,t_0)$ (1.1) that since then has served as the basis for the optimal portfolio selections and for the development of investment strategies. Markowitz's expression of the variance is valid for any portfolio and for the market portfolio as well. We follow Markowitz's notations, and $X_j(t_0)$ (2.3) defines the relative amount invested into security j of the portfolio at time $t_0$ in the past. Functions $\theta_{jk}(t,t_0)$ (B.9) denote the covariances between returns of securities *j* and *k*.

$$\Theta_M(t, t_0) = \sum_{j,k=1}^{J} \theta_{jk}(t, t_0) \; X_j(t_0) X_k(t_0) \qquad (1.1)$$

Meanwhile, the thesis that "the devil is in the details" is still actual. Olkhov (2025a-2025c) showed that the expression of the variance $\Theta_M(t,t_0)$ (1.1) that was derived by Markowitz (1952) implicitly describes a rather simplified approximation of the market trades with the securities of the portfolio when the volumes of all consecutive trades with each security are assumed constant.

However, real financial markets reveal highly irregular or random variations of the volumes of consecutive trades. The approximation that accounts for the effect of random variations of the trade volumes describes the market-based variance of the portfolio. The distinctions between Markowitz's approximation of the variance $\Theta_M(t,t_0)$ (1.1), which ignores the random variations of the trade volumes, and the market-based variance that accounts for the effect of the random variations of the volumes of consecutive trades can be sufficiently high (Olkhov, 2025b).

In this paper we consider the investor who doesn't trade the shares of his market portfolio that he composed at time $t_0$ in the past. At current time *t*, the investor observes the time series of the trades made with all securities in the market during a particular averaging interval and converts these time series into the time series, which model the trades with his market portfolio as trades with a single security. We present the unified description of the market-based variances for the trades with particular securities, for the trades with all market securities as trades with a single security, and the market-based variance of the market portfolio. The investor, who observes the trades with all securities in the market, obtains the current estimates of the return and variance of his unchanged portfolio.

We describe the dependence between the return and the variance of trades with all securities in the market and the return and the variance of the market portfolio. We present some cases with different variations of the volumes of consecutive trades with all securities in the market and with the market portfolio. We compare Markowitz's approximations of the variances of all trades in the market and of the market portfolio and show possible distinctions between them. We use the Taylor series to highlight the impact of random variations of the volumes of consecutive trades and derive the dependence of the market portfolio variance on the coefficients of variation of the random values and volumes of consecutive trades. Finally, we discuss the importance of the averaging interval for the assessment of market-based variances of the portfolios of different value and outline the obstacles that prevent reliable forecasting of returns and variances with high accuracy.

## 2. Main notations

In this section we introduce the principal notations that help describe the entire market composed of shares of all securities traded in the market. Then, we introduce the notations that describe the market portfolio that was collected by the investor at time $t_0$ in the past as a slice of the entire market. Further, at current time $t$, we consider the time series of market trades with the shares of all securities in the market during a particular time averaging interval.

### 2.1 The entire market at time $t_0$ in the past

Let us assume that the entire market is composed of shares of $J$ securities traded in the market. For simplicity we assume that since time $t_0$ in the past, the number or the volume $W_{mj}(t_0)$ of shares outstanding of each security $j=1,...J$ in the market is constant. The market capitalization, or the total value $Q_{mj}(t_0)$ (2.1), of shares of security $j$ at time $t_0$ is determined by its price $p_j(t_0)$:

$$Q_{mj}(t_0) = p_j(t_0) W_{mj}(t_0) \qquad (2.1)$$

The capitalization of the entire market, or the total market value $Q_m(t_0)$ (2.2) of all securities $j=1,...J$ traded in the market at time $t_0$, is a sum of their market capitalization $Q_{mj}(t_0)$ (2.1). The total number of shares outstanding, $W_m(t_0)$ (2.2), is a sum of shares of all securities and is a constant. We introduce the market price $s_m(t_0)$ (2.2) per share outstanding of all securities:

$$Q_m(t_0) = \sum_{j=1}^{J} Q_{mj}(t_0) = \sum_{j=1}^{J} p_j(t_0) W_{mj}(t_0) \; ; \; W_m(t_0) = \sum_{j=1}^{J} W_{mj}(t_0); \; Q_m(t_0) = s_m(t_0) W_m(t_0) \qquad (2.2)$$

We follow Markowitz (1952), and note $X_j(t_0)$ (2.3) the relative capitalization of security $j$ in the entire market at time $t_0$:

$$X_j(t_0) = \frac{Q_{mj}(t_0)}{Q_m(t_0)} = \frac{p_j(t_0)W_{mj}(t_0)}{s_m(t_0)\cdot W_m(t_0)} = \frac{p_j(t_0)}{s_m(t_0)} x_j(t_0) \; ; \; x_j(t_0) = \frac{W_{mj}(t_0)}{W_m(t_0)} \; ; \; \sum_{j=1}^{J} x_j(t_0) = \sum_{j=1}^{J} X_j(t_0) = 1 \quad (2.3)$$

The functions $x_j(t_0)$ (2.3) denote the relative number of the shares $W_{mj}(t_0)$ (2.1) of security $j$ in the total number of shares $W_m(t_0)$ (2.2) of the entire market. The functions $Q_{mj}(t_0)$, $W_{mj}(t_0)$, and $p_j(t_0)$ (2.1) describe the values, volumes, and prices of securities $j=1,...J$ in the entire market at time $t_0$. The functions $x_j(t_0)$ and $X_j(t_0)$ (2.3) describe the relative number of shares and the relative capitalization of securities $j$ in the entire market. The function $Q_m(t_0)$ (2.2) describes the total value or total capitalization of the entire market at time $t_0$ in the past, with the total volume $W_m(t_0)$ (2.2) of shares of all market securities and the price $s_m(t_0)$ (2.2) per share at time $t_0$.

## 2.2 The market portfolio at time $t_0$ in the past

Let us assume that at time $t_0$, the investor collected his portfolio of the total value $Q_\Sigma(t_0)$ as a piece of the entire market. The investor collected the shares of all securities $j=1,..J$ that are traded in the market. The investor purchased $W_j(t_0)$ (2.4) shares of security $j$ at price $p_j(t_0)$ (2.1). The value $Q_j(t_0)$ of shares of security $j$ in his portfolio is proportional to the market shares $X_j(t_0)$ (2.3) of their capitalization. The value $Q_j(t_0)$ and volume $W_j(t_0)$ obey (2.4; 2.5):

$$Q_j(t_0) = Q_\Sigma(t_0) \cdot X_j(t_0) = p_j(t_0) \cdot W_j(t_0) \quad ; \quad X_j(t_0) = \frac{Q_{mj}(t_0)}{Q_m(t_0)} = \frac{Q_j(t_0)}{Q_\Sigma(t_0)} \quad (2.4)$$

The number $W_j(t_0)$ (2.5) of shares of security $j$ in the market portfolio is proportional to the number $W_{mj}(t_0)$ of shares of security $j$ in the entire market (2.5):

$$W_j(t_0) = \frac{Q_\Sigma(t_0)}{Q_m(t_0)} W_{mj}(t_0) \quad ; \quad Q_j(t_0) = p_j(t_0) \frac{Q_\Sigma(t_0)}{Q_m(t_0)} W_{mj}(t_0) \quad (2.5)$$

The total value $Q_\Sigma(t_0)$ (2.6) and total volume $W_\Sigma(t_0)$ (2.6) of the market portfolio at time $t_0$:

$$Q_\Sigma(t_0) = \sum_{j=1}^{J} Q_j(t_0) \; ; \; W_\Sigma(t_0) = \sum_{j=1}^{J} W_j(t_0) = \frac{Q_\Sigma(t_0)}{Q_m(t_0)} W_m(t_0) \; ; \; x_j(t_0) = \frac{W_j(t_0)}{W_\Sigma(t_0)} = \frac{W_{mj}(t_0)}{W_m(t_0)} \quad (2.6)$$

The relative number $x_j(t_0)$ (2.6) of shares of security $j$ in the market portfolio equals the relative number of shares $x_j(t_0)$ (2.3) in the entire market. The prices per share $p_j(t_0)$ of securities $j$ are the same as same prices in the entire market. The price per share $s(t_0)$ (2.7) of the market portfolio is the same as the price $s_m(t_0)$ (2.2) per share of the entire market.

$$Q_\Sigma(t_0) = s(t_0) \cdot W_\Sigma(t_0) \quad ; \quad Q_m(t_0) = s_m(t_0) \cdot W_m(t_0) \quad ; \quad s(t_0) = s_m(t_0) \quad (2.7)$$

Indeed, from (2.4-2.6), obtain

$$W_\Sigma(t_0) = \sum_{j=1}^{J} W_j(t_0) = \sum_{j=1}^{J} \frac{Q_\Sigma(t_0)}{Q_m(t_0)} W_{mj}(t_0) = \frac{Q_\Sigma(t_0)}{Q_m(t_0)} W_m(t_0) \to s(t_0) = \frac{Q_\Sigma(t_0)}{W_\Sigma(t_0)} = \frac{Q_m(t_0)}{W_m(t_0)} = s_m(t_0) \quad (2.8)$$

We highlight that the price $s(t_0)$ (2.9) per share of the market portfolio has the form of volume weighted average price (VWAP) (Berkowitz et al., 1988) by the volumes of shares of securities $j=1,...J$ of the portfolio or by the volumes of shares of the entire market:

$$s(t_0) = \frac{Q_\Sigma(t_0)}{W_\Sigma(t_0)} = \frac{1}{W_\Sigma(t_0)}\sum_{j=1}^{J} p_j(t_0)\, W_j(t_0) = \frac{1}{W_m(t_0)}\sum_{j=1}^{J} p_j(t_0)\, W_{mj}(t_0) = \frac{Q_m(t_0)}{W_m(t_0)} = s_m(t_0) \quad (2.9)$$

We assume that the investor doesn't trade his shares and the number of shares $W_j(t_0)$ of each security $j=1,...J$ in his portfolio is constant. However, the investor may be interested in current assessments of returns, variance, and risks of his portfolio. To estimate them at current time $t$, the investor may observe the time series of the market trades with all $j=1,...J$ securities during a particular averaging interval.

For simplicity we assume that market trades with all securities are made simultaneously at time $t_i$ with a short span $\varepsilon>0$ between consecutive trades. We assume that $\varepsilon$ is constant and the same for all securities. The investor chooses the time averaging interval $\varDelta$ (2.10) that contains the $N$ terms of time series $t_i$ of trades with each security $j=1,...J$ of the entire market:

$$\Delta = \left[t - \frac{\Delta}{2};\; t + \frac{\Delta}{2}\right] \quad;\quad t_i \in \Delta \quad;\quad i = 1,\ldots N \quad;\quad N \cdot \varepsilon = \Delta \quad (2.10)$$

### *2.3 The market trades with securities $j=1,...J$ at current time t*

We denote the value $C_j(t_i)$ and the volume $U_j(t_i)$ of trade with shares of security $j$ at time $t_i$ during $\varDelta$ (2.10). Each trade with security $j$ at time $t_i$ determines its current price $p_j(t_i)$ (2.11):

$$C_j(t_i) = p_j(t_i) U_j(t_i) \quad;\quad j = 1,\ldots J \quad (2.11)$$

We define the mean $C(t;1)$, $U(t;1)$ values and volumes and the mean squares $C(t;2)$, $U(t;2)$ (2.12) of the values and volumes of trades during $\varDelta$:

$$C_j(t;n) = \frac{1}{N}\sum_{i=1}^{N} C_j^n(t_i) \quad;\quad U_j(t;n) = \frac{1}{N}\sum_{i=1}^{N} U_j^n(t_i) \quad;\quad n = 1,2. \quad (2.12)$$

The total values $C_{\Sigma j}(t)$ and volumes $U_{\Sigma j}(t)$ (2.13) of trades with security $j$ during $\varDelta$ (2.10)

$$C_{\Sigma j}(t) = \sum_{i=1}^{N} C_j(t_i) = N \cdot C_j(t;1) \quad;\quad U_{\Sigma j}(t) = \sum_{i=1}^{N} U_j(t_i) = N \cdot U_j(t;1) \quad (2.13)$$

The equations (2.11-2.13) define the mean prices $p_j(t;1)$ (2.14) of securities $j=1,...J$ during $\varDelta$:

$$p_j(t;1) = \frac{C_{\Sigma j}(t)}{U_{\Sigma j}(t)} = \frac{1}{U_{\Sigma j}(t)}\sum_{i=1}^{N} p_j(t_i)\, U_j(t_i) = \frac{C_j(t;1)}{U_j(t;1)} \quad (2.14)$$

The mean price $p_j(t;1)$ (2.14) has the form of VWAP.

### 3. The time series of trades with the market portfolio as with a single security

We recall that the investor doesn't trade the shares of his market portfolio. The investor observes the time series (2.11; 2.12) of trades with all market securities during the averaging interval $\varDelta$ (2.10) and transforms them into the time series that model the trades with his portfolio as trades with a single security. To show how to do that, let us for each security *j* determine the scale $\lambda_j$ (3.1) that equals the ratio of the number $W_j(t_0)$ (2.5) of shares of security *j* in the market portfolio at time $t_0$ to the total volume $U_{\Sigma j}(t)$ (2.13) of trades with security *j* in the market at current time *t* during $\Delta$:

$$\lambda_j = \frac{W_j(t_0)}{U_{\Sigma j}(t)} \quad ; \quad c_j(t_i) = \lambda_j \cdot C_j(t_i) \quad ; \quad u_j(t_i) = \lambda_j \cdot U_j(t_i) \qquad (3.1)$$

The equations (3.1) introduce the normalized values $c_j(t_i)$ and volumes $u_j(t_i)$ of trades with securities *j=1,...J* during $\Delta$. It is evident that the change of variables (3.1) doesn't change the prices $p_j(t_i)$ in equations (2.11) and (3.2):

$$c_j(t_i) = p_j(t_i)\, u_j(t_i) \quad or \quad \lambda_j \cdot C_j(t_i) = p_j(t_i) \cdot \lambda_j \cdot U_j(t_i) \qquad (3.2)$$

From (2.13; 3.1), obtain that the total normalized volume $u_{\Sigma j}(t)$ (3.3) of trades with security *j* during $\Delta$ equals the number of shares $W_j(t_0)$ (2.5) of security *j* in the market portfolio at time $t_0$:

$$u_{\Sigma j}(t) = \sum_{i=1}^{N} u_j(t_i) = \frac{W_j(t_0)}{U_{\Sigma j}(t)} \cdot \sum_{i=1}^{N} U_j(t_i) = W_j(t_0) \qquad (3.3)$$

For each security *j=1,...J*, we consider the time series of the normalized values $c_j(t_i)$ and volumes $u_j(t_i)$ (3.1) of trades as a model of trades with the total trade volumes $u_{\Sigma j}(t)$ (3.3) during $\Delta$ that are exactly equal to the numbers of shares $W_j(t_0)$ (2.5) of security *j=1,...J* in the market portfolio. The sums (3.4) of the normalized values $c_j(t_i)$ and volumes $u_j(t_i)$ (3.1) of trades with all securities *j=1,...J* at time $t_i$ determine the time series of the values $Q(t_i)$ and volumes $W(t_i)$ (3.4) of trades with the portfolio as trades with a single security at price $s(t_i)$ (3.4):

$$Q(t_i) = \sum_{j=1}^{J} c_j(t_i) \quad ; \quad W(t_i) = \sum_{j=1}^{J} u_j(t_i) \quad ; \quad Q(t_i) = s(t_i) \cdot W(t_i) \qquad (3.4)$$

We recall that the investor doesn't trade shares of his market portfolio, and the time series of the values $Q(t_i)$ and volumes $W(t_i)$ (3.4) only model the trades with the portfolio. The total volume $W_\Sigma(t)$ (3.5) of trades equals the total number $W_\Sigma(t_0)$ (2.6) of shares in the market portfolio. The total value $Q_\Sigma(t)$ (3.5) of trades during $\Delta$ estimates the current value of the portfolio:

$$W_\Sigma(t) = \sum_{i=1}^{N} W(t_i) = W_\Sigma(t_0) \quad ; \quad Q_\Sigma(t) = \sum_{i=1}^{N} Q(t_i) \quad ; \quad Q_\Sigma(t) = s(t) \cdot W_\Sigma(t) \qquad (3.5)$$

The total value $Q_\Sigma(t)$ and volume $W_\Sigma(t)$ (3.5) of trades at current time *t* define the mean price $s(t;1)$ (3.5;3.6) per share of the investor's market portfolio during $\Delta$ as VWAP.

$$s(t;1) = \frac{Q_\Sigma(t)}{W_\Sigma(t)} = \frac{1}{W_\Sigma(t_0)} \sum_{i=1}^{N} s(t_i) \cdot W(t_i) = \frac{1}{W(t;1)} \frac{1}{N} \sum_{i=1}^{N} s(t_i) \cdot W(t_i) = \frac{Q(t;1)}{W(t;1)} \qquad (3.6)$$

Functions $Q(t;1)$ and $W(t;1)$ in (3.6) denote the mean value and volume during $\Delta$ (2.10):

$$Q(t;1) = \frac{1}{N}\sum_{i=1}^{N} Q(t_i) \quad ; \quad W(t;1) = \frac{1}{N}\sum_{i=1}^{N} W(t_i) \tag{3.7}$$

## 4. The time series trades with the entire market as trades with a single security

Now we consider the time series of the values $C_j(t_i)$ and the volumes $U_j(t_i)$ (2.11-2.13) of trades at time $t_i$ with all market securities $j=1,...J$ and transform them alike as we did to derive the time series of the values $Q(t_i)$ and volumes $W(t_i)$ (3.4) that model the trades with the market portfolio as the trades with a single security. To transform the time series of the values $C_j(t_i)$ and the volumes $U_j(t_i)$ (2.11-2.13) of trades with securities $j=1,...J$ at time $t_i$ during $\Delta$ (2.10) and to obtain the time series of all market trades as trades with a single security, let us sum (4.1) at time $t_i$ the values $C_j(t_i)$ and volumes $U_j(t_i)$ of trades with all securities $j=1,...J$ in the market during $\Delta$:

$$C_m(t_i) = \sum_{j=1}^{J} C_j(t_i) \quad ; \quad U_m(t_i) = \sum_{j=1}^{J} U_j(t_i) \quad ; \quad C_m(t_i) = s_m(t_i) \cdot U_m(t_i) \tag{4.1}$$

The values $C_m(t_i)$ and volumes $U_m(t_i)$ (4.1) at time $t_i$ determine the price $s_m(t_i)$ (4.1) per share of all securities in the market. The total values $C_{\Sigma m}(t)$ and volumes $U_{\Sigma m}(t)$ (4.2) of all trades during $\Delta$ take the form:

$$C_{\Sigma m}(t) = \sum_{i=1}^{N} C_m(t_i) = \sum_{j=1}^{J} C_{\Sigma j}(t) \quad ; \quad U_{\Sigma m}(t) = \sum_{i=1}^{N} U_m(t_i) = \sum_{j=1}^{J} U_{\Sigma j}(t) \tag{4.2}$$

Functions $C_{\Sigma j}(t)$ and $U_{\Sigma j}(t)$ in (2.13; 4.2) define the total values and volumes of trades with security $j$ during $\Delta$. The total values $C_{\Sigma m}(t)$ and volumes $U_{\Sigma m}(t)$ (4.2) define the mean price $s_m(t;1)$ (4.3) per share of all securities in the entire market during $\Delta$ as VWAP:

$$C_{\Sigma m}(t) = s_m(t) \cdot U_{\Sigma m}(t) \quad ; \quad s_m(t;1) = \frac{C_{\Sigma m}(t)}{U_{\Sigma m}(t)} = \frac{1}{U_{\Sigma m}(t)} \sum_{i=1}^{N} s_m(t_i) \cdot U_m(t_i) \tag{4.3}$$

At current time $t$, we define $X_j(t)$ (4.4) as the ratio of the total values $C_{\Sigma j}(t)$ (2.13) of trades with security $j$ to the total value $C_{\Sigma m}(t)$ (4.3) of all market trades during $\Delta$:

$$X_j(t) = \frac{C_{\Sigma j}(t)}{C_{\Sigma m}(t)} \quad ; \quad x_j(t) = \frac{U_{\Sigma j}(t)}{U_{\Sigma m}(t)} \quad ; \quad \sum_{j=1}^{J} X_j(t) = \sum_{j=1}^{J} x_j(t) = 1 \tag{4.4}$$

The relative values $X_j(t)$ and the relative volumes $x_j(t)$ (4.4) of the current market trades with security $j$ can be different from the relative amounts $X_j(t_0)$ (2.3) invested into security $j$ and the relative volumes $x_j(t_0)$ (2.3) of shares in the portfolio at time $t_0$ in the past. We recall that $X_j(t_0)$ and $x_j(t_0)$ (2.3) determine the composition (2.4; 2.5) of the investor's market portfolio at time $t_0$. The proportions of the values and volumes of market trades with the securities at the current time $t$ during $\Delta$ can be different from the proportions of the values and volumes of securities in the investor's market portfolio at time $t_0$.

The mean price $s(t;1)$ (3.5; 3.6) per share of the market portfolio can be different from the mean price $s_m(t;1)$ (4.3) per share of all market trades as trades with a single security. Indeed, at the current time $t$, the time series of the values $C_m(t_i)$ and the volumes $U_m(t_i)$ (4.1) of all market trades as trades with a single security determine the relative values $X_j(t)$ (4.4) of trades with security $j$ during $\Delta$ that are different from the relative amounts $X_j(t_0)$ (2.3) of the capitalization of security $j$ on the total market that are equal to the relative amount invested into security $j$ in the market portfolio. Meanwhile, the time series of the values $Q(t_i)$ and volumes $W(t_i)$ (3.4) that model the trades with the portfolio as with a single security preserve its relative number of shares $x_j(t_0)$ and the relative amounts $X_j(t_0)$ (2.3) invested into securities at time $t_0$ in the past. That explains the origin of the distinctions at current time $t$ between the mean price per share $s(t;1)$ (3.5; 3.6) of the investor's market portfolio and the mean price per share $s_m(t;1)$ (4.3) of all market trades as trades with a single security. What are the consequences?

## 5. Random returns, mean returns, and variances

The time series of the values $C_j(t_i)$ and the volumes $U_j(t_i)$ (2.11-2.14) of trades with securities, the time series of the values $Q(t_i)$ and volumes $W(t_i)$ (3.4-3.6) that model the trades with the portfolio as with a single security, and the time series of $C_m(t_i)$ and $U_m(t_i)$ (4.1-4.3) of all market trades as trades with a single security determine the corresponding random and mean returns and variances.

### 5.1 Random returns, mean return, and the variance of security $j$

We define the random $R_j(t_i,t_0)$ returns and the mean return $R_j(t,t_0)$ (5.1) of security $j$ as a result of a trade (2.11) at time $t_i$ during $\Delta$ (2.10) with respect to the price $p_j(t_0)$ of security $j$ at time $t_0$:

$$R_j(t_i, t_0) = \frac{p_j(t_i)}{p_j(t_0)} \quad ; \quad R_j(t, t_0) = \frac{p_j(t;1)}{p_j(t_0)} = \frac{C_{\Sigma j}(t)}{p_j(t_0) U_{\Sigma j}(t)} = \frac{1}{U_{\Sigma j}(t)} \sum_{i=1}^{N} R_j(t_i, t_0) U_j(t_i) \quad (5.1)$$

We use "gross" return (5.1) instead of the more usual definition of return $r_j(t_i,t_0) = R_j(t_i,t_0)-1$, but the variances of both definitions are the same. The VWAP $p_j(t;1)$ (2.14; 5.1) determines the volume weighted average return $R_j(t,t_0)$ (5.1) of trades with security $j$. The market-based variance $\Theta_j(t,t_0)$ (5.2) that accounts for the impact of the random variations of the volumes $U_j(t_i)$ of consecutive trades with security $j$ takes the form:

$$\Theta_j(t, t_0) = \frac{\psi_j^2 - 2\varphi_j + \chi_j^2}{1 + \chi_j^2} R_j^2(t, t_0) \quad (5.2)$$

In App. A (A.8), we give a brief derivation of (5.2). The variance $\Theta_j(t,t_0)$ (5.2) depends upon the coefficients of variation $\psi_j$ and $\chi_j$ (5.3) of the values $C_j(t_i)$ and volumes $U_j(t_i)$ of trades during $\Delta$:

$$\psi_j{}^2 = \frac{cov\{C_j(t),C_j(t)\}}{C_j^2(t;1)} \quad ; \quad \chi_j{}^2 = \frac{cov\{U_j(t),U_j(t)\}}{U_j^2(t;1)} \qquad (5.3)$$

The covariances $cov\{C_j(t),C_j(t)\}$ and $cov\{U_j(t),U_j(t)\}$ of values and volumes have the usual form:

$$cov\{C_j(t),C_j(t)\} = \tfrac{1}{N}\sum_{i=1}^{N}\left(C_j(t_i) - C_j(t;1)\right)^2 \quad ; \quad cov\{U_j(t),U_j(t)\} = \tfrac{1}{N}\sum_{i=1}^{N}\left(U_j(t_i) - U_j(t;1)\right)^2 \qquad (5.4)$$

For simplicity, we omit the dependence of the coefficients of variation $\psi_j$ and $\chi_j$ (5.3) and the function $\varphi_j$ (5.5) on time $t$. The function $\varphi_j$ (5.5) denotes the ratio of the covariance of the values $C_j(t_i)$ and the volumes $U_j(t_i)$ of trades with security $j$ to their mean values.

$$\varphi_j = \frac{cov\{C_j(t),U_j(t)\}}{C_j(t;1)U_j(t;1)} \quad ; \quad cov\{C_j(t),U_j(t)\} = \tfrac{1}{N}\sum_{i=1}^{N}\left(C_j(t_i) - C_j(t;1)\right)\left(U_j(t_i) - U_j(t;1)\right) \qquad (5.5)$$

The mean square $U_j(t;2)$ (2.12) of the time series of trade volumes $U_j(t_i)$ takes the form (5.6):

$$U_j(t;2) = \tfrac{1}{N}\sum_{i=1}^{N} U_j^2(t_i) = U_j^2(t;1)[1 + \chi_j{}^2] \qquad (5.6)$$

The time series of the values $C_j(t_i)$ and the volumes $U_j(t_i)$ of consecutive trades with securities $j=1,...J$ determine their random $R_j(t_i,t_0)$ returns, the mean returns $R_j(t,t_0)$ (4.1), and the variances $\Theta_j(t,t_0)$ (5.2) during the time averaging interval $\Delta$.

## 5.2 Random returns, mean return, and the variance of the current market trades

The substitution of the time series of the values $C_m(t_i)$ and volumes $U_m(t_i)$ (4.1-4.3) that describe the current trades with all securities as the trades with a single security, instead of the time series of the values $C_j(t_i)$ and volumes $U_j(t_i)$ (2.11-2.14) of trades with a particular market security $j$, defines the random return $R_m(t_i,t_0)$ and the mean return $R_m(t,t_0)$ (5.7) of the current trades as (5.1):

$$R_m(t_i,t_0) = \frac{s_m(t_i)}{s_m(t_0)} \quad ; \quad R_m(t,t_0) = \frac{s_m(t;1)}{s_m(t_0)} = \frac{C_{\Sigma m}(t)}{s_m(t_0)U_{\Sigma m}(t)} = \frac{1}{U_{\Sigma m}(t)}\sum_{i=1}^{N} R_m(t_i,t_0) U_m(t_i) \qquad (5.7)$$

The market-based variance $\Theta_m(t,t_0)$ (5.8) that accounts for the effects of the random variations of the volumes $U_m(t_i)$ (4.1-4.3) of the consecutive trades as trades with a single security during $\Delta$ has the same form as the variance $\Theta_j(t,t_0)$ (5.2) of security $j$:

$$\Theta_m(t,t_0) = \frac{\psi_m{}^2 - 2\varphi_m + \chi_m{}^2}{1+\chi_m{}^2} R_m^2(t,t_0) \qquad (5.8)$$

The substitutions of the time series of $C_m(t_i)$ and $U_m(t_i)$ (4.1-4.3) instead of the time series of $C_j(t_i)$ and $U_j(t_i)$ (2.11-2.14) define the coefficients of variation $\psi_m$ and $\chi_m$ of the values $C_m(t_i)$ and volumes $U_m(t_i)$ of all market trades as trades with a single security (5.9):

$$\psi_m{}^2 = \frac{cov\{C_m(t),C_m(t)\}}{C_m^2(t;1)} \quad ; \quad \chi_m{}^2 = \frac{cov\{U_m(t),U_m(t)\}}{U_m^2(t;1)} \quad ; \quad \varphi_m = \frac{cov\{C_m(t),U_m(t)\}}{C_m(t;1)U_m(t;1)} \qquad (5.9)$$

## 5.3 Random returns, mean return, and the variance of the market portfolio

The time series of the values $Q(t_i)$ and volumes $W(t_i)$ (3.4; 3.5) that model the trades with the market portfolio of the investor as the trades with a single security give the same unified estimates of the random $R(t_i,t_0)$ returns, the mean return $R(t,t_0)$ (5.10), and the variance $\Theta(t,t_0)$ (5.11) of the market portfolio at current time $t$ during $\Delta$.

$$R(t_i, t_0) = \frac{s(t_i)}{s(t_0)} \quad ; \quad R(t, t_0) = \frac{s(t;1)}{s(t_0)} = \frac{Q_\Sigma(t)}{s(t_0)W_\Sigma(t)} = \frac{1}{W_\Sigma(t)} \sum_{i=1}^{N} R(t_i, t_0) W(t_i) \quad (5.10)$$

We recall that the investor doesn't trade the shares of his market portfolio and (5.10-5.11) only model the return and the variance of the market portfolio using the time series of current trades with all market securities during $\Delta$. The market-based variance $\Theta(t,t_0)$ (5.11; A.8) of the market portfolio that accounts for the effects of the random fluctuations of the volumes $W(t_i)$ (3.4) of the consecutive trades with the portfolio takes the form:

$$\Theta(t, t_0) = \frac{\psi^2 - 2\varphi + \chi^2}{1 + \chi^2} R^2(t, t_0) \quad (5.11)$$

The estimates (5.10; 5.11) are the results of the current observations of the time series of trades with all securities in the market during $\Delta$ (2.10). The coefficients of variation $\psi$ and $\chi$ (5.12) of the values $Q(t_i)$ and volumes $W(t_i)$ of trades with the market portfolio, as with a single security, take the form:

$$\psi^2 = \frac{cov\{Q(t),Q(t)\}}{Q^2(t;1)} \quad ; \quad \chi^2 = \frac{cov\{W(t),W(t)\}}{W^2(t;1)} \quad ; \quad \varphi = \frac{cov\{Q(t),W(t)\}}{Q(t;1)W(t;1)} \quad (5.12)$$

We highlight that at current time $t$ during $\Delta$ the equations (5.1-5.3), (5.7-5.9), and (5.10-5.12) give the unified description of the random returns, the mean returns, and the variances of each market security $j=1,...J$, of the market trades as trades with a single security, and of the investor's market portfolio that was composed at time $t_0$ and since then remains unchanged. Let us consider the distinctions between these assessments.

## 6. Investors' dreams and market reality

One of the main dreams of the investors who hold their market portfolios collected at time $t_0$ in the past is to gain at current time $t$ the return under the variance and the risks of the trade with the entire market. Market reality may correct their dreams.

The difference between the current estimate of the mean price $s(t;1)$ (3.5) per share of the investor's market portfolio and the mean price $s_m(t;1)$ (4.3) per share of the entire market trade as trade with a single security during $\Delta$ reveals that the mean return and the variance of the investor's market portfolio (5.10; 5.11) may differ from the return and the variance (5.7; 5.8) of the entire market trades. The mean return $R_m(t,t_0)$ (5.7) of all market trades and the estimate of the mean return $R(t,t_0)$ (5.10) of the market portfolio depend on the mean prices $s_m(t;1)$ (4.3) and $s(t;1)$ (3.5; 3.6). From (2.17), obtain relations (6.1) on the mean price $s_m(t;1)$:

$$s_m(t;1) = \sum_{j=1}^{J} \frac{U_{\Sigma j}(t)}{U_{\Sigma m}(t)} \frac{1}{U_{\Sigma j}(t)} \sum_{i=1}^{N} p_j(t_i) \cdot U_j(t_i) = \sum_{j=1}^{J} p_j(t;1) \cdot x_j(t) \; ; \; x_j(t) = \frac{U_{\Sigma j}(t)}{U_{\Sigma m}(t)} \quad (6.1)$$

We recall that $x_j(t)$ (4.4; 6.1) denotes the ratio of the volumes $U_{\Sigma j}(t)$ (2.13) of trades with security $j$ to the total volume $U_{\Sigma m}(t)$ (4.3) of all market trades during $\Delta$. The similar expression for the mean price $s(t)$ (3.5; 3.6) of the market portfolio takes the form:

$$s(t;1) = \sum_{j=1}^{J} \frac{1}{W_\Sigma(t)} \frac{W_j(t_0)}{U_{\Sigma j}(t)} \sum_{i=1}^{N} p_j(t_i) \cdot U_j(t_i) = \sum_{j=1}^{J} p_j(t;1) \cdot x_j(t_0) \; ; \; x_j(t_0) = \frac{W_j(t_0)}{W_\Sigma(t_0)} \quad (6.2)$$

$x_j(t_0)$ denotes the ratio of the relative number of shares $W_j(t_0)$ (2.4; 2.5) of security $j$ to the total number $W_\Sigma(t_0)$ (2.6) of shares of the portfolio. The difference between (6.2) and (6.1) gives:

$$s(t;1) - s_m(t;1) = \sum_{j=1}^{J} p_j(t;1) \, x_j(t_0) \left[1 - \frac{x_j(t)}{x_j(t_0)}\right] \quad (6.3)$$

### 6.1 The mean return of the market portfolio and the mean return of all market trades

From (5.7; 5.10) and (2.4), after simple transformations, obtain the relations for the mean returns:

$$R_m(t,t_0) = \sum_{j=1}^{J} R_j(t,t_0) \cdot X_j(t_0) \cdot \frac{x_j(t)}{x_j(t_0)} \quad ; \quad R(t,t_0) = \sum_{j=1}^{J} R_j(t,t_0) \cdot X_j(t_0) \quad (6.4)$$

One can mention that the mean $R(t,t_0)$ (6.4) of the investor's portfolio, as it should be, coincides with the expression of the mean return that was derived by Markowitz (1952). The relative amount $X_j(t_0)$ (4.4) takes the form (2.3; 6.5):

$$X_j(t_0) = \frac{Q_j(t_0)}{Q_\Sigma(t_0)} = \frac{p_j(t_0)}{s(t_0)} \cdot x_j(t_0) = \frac{Q_{mj}(t_0)}{Q_m(t_0)} \quad (6.5)$$

Finally, from (6.4; 6.5), obtain the difference between the mean return $R(t,t_0)$ (5.10) of the market portfolio and the mean return $R_m(t,t_0)$ (5.7; 6.4) of all market trades:

$$R(t,t_0) - R_m(t,t_0) = \sum_{j=1}^{J} R_j(t,t_0) \cdot X_j(t_0) \cdot \left[1 - \frac{x_j(t)}{x_j(t_0)}\right] \quad (6.6)$$

The use of (6.3) give another form of (6.6):

$$R(t,t_0) - R_m(t,t_0) = \frac{s(t;1) - s_m(t;1)}{s_m(t_0)} = \frac{1}{s_m(t_0)} \sum_{j=1}^{J} p_j(t;1) \, x_j(t_0) \left[1 - \frac{x_j(t)}{x_j(t_0)}\right] \quad (6.7)$$

If the relative volumes $x_j(t)$ (6.1) of current trades with security $j$ in the total volume of all market trades equal the relative number of shares $x_j(t_0)$ (6.2) of security $j$ in the market portfolio of the investor, the mean return $R(t,t_0)$ (6.4) of the investor's portfolio equals the mean return $R_m(t,t_0)$ (5.7; 6.4) of market trades at current time $t$ during $\Delta$. However, the current relative volumes $x_j(t)$ (6.1) of market trades during $\Delta$ can be different from the relative number of shares $x_j(t_0)$ (6.2) of the entire market and the relative number $x_j(t_0)$ of shares of the market portfolio at time $t_0$.

That highlights the economic origin of the distinctions between the performance of the entire market and the market portfolio on one hand, and the performance of the current market trades with all securities during $\Delta$ on the other hand. The rise of the relative volumes $x_j(t)$ (6.1) of current trades to compare with the relative number of shares $x_j(t_0)$ (6.2) of security $j$ in the entire market increases the current contribution of the mean returns $R_j(t,t_0)$ (5.1) of trades with security $j$ into the mean return $R_m(t,t_0)$ (5.7; 6.4) of all market trades during $\Delta$. The decline of $x_j(t)$ (6.1) downturns the contribution of the mean returns $R_j(t,t_0)$ (5.1) of security $j$ into the mean return $R_m(t,t_0)$ (5.7; 6.4) of all market trades with respect to the mean return $R(t,t_0)$ (6.4) of the investor's portfolio. The market portfolio preserves constant the relative amounts $X_j(t_0)$ (6.5) invested into securities $j=1,...J$ at time $t_0$, as they were determined by the entire market. The relative volumes $x_j(t)$ (6.1) of all market trades at current time $t$ during $\Delta$ change the factors $X_j(t_0)x_j(t)/x_j(t_0)$ that determine the contribution of the mean returns $R_j(t,t_0)$ (5.1) of security $j$ into the mean return $R_m(t,t_0)$ (6.4) of all market trades.

We highlight that at the current time $t$ the trade with all market securities $j=1,...J$ has the mean return $R_m(t,t_0)$ (6.4) with respect to the price $s_m(t_0)$ (2.2) per share of the entire market at time $t_0$ in the past, which differs from the current estimates of the mean return $R(t,t_0)$ (6.4) of the market portfolio with respect to the same price $s_m(t_0)$ at $t_0$.

The return of current trade with all market securities $j=1,...J$ is different from the return of the market portfolio. The investors and the portfolio managers who hold market portfolios should keep that in mind and be ready to meet such distinctions.

### *6.2 The variance of the current market trade and the variance of the market portfolio*

The expressions of the market-based variance $\Theta_m(t,t_0)$ (4.8) of all market trades and the market-based variance $\Theta(t,t_0)$ (4.11) of the market portfolio that account for the effects of the random variations of the volumes of consecutive trades are rather difficult for their direct comparison. To simplify their comparisons, one may use their Taylor expansions up to the 2nd degree by the coefficients of variation of the volumes of consecutive trades during $\Delta$ (Olkhov, 2025b). The Taylor series of the market-based variance $\Theta(t,t_0)$ (5.11) of the market portfolio takes the form:

$$\Theta(t,t_0) = \Theta_M(t,t_0) - 2a\,\Theta_M^{1/2}(t,t_0)\,R(t,t_0)\,\chi + [R^2(t,t_0) - \Theta_M(t,t_0)]\,\chi^2 \qquad (6.8)$$

*R(t,t₀)* (5.10; 6.4) denotes the mean return of the market portfolio. The function *Θ*$_M$*(t,t₀)* in (6.8) denotes Markowitz variance (6.9). Olkhov (2025a-2025c) has shown that the famous Markowitz (1952) expression of the portfolio variance *Θ*$_M$*(t,t₀)* (6.9) describes only a rather simplified approximation of the real markets when all volumes $U_j(t_i)$ of consecutive trades with securities *j=1,..J* are assumed to be constant during the averaging interval *Δ* (2.10).

$$\Theta_M(t,t_0) = \sum_{j,k=1}^{J} \theta_{jk}(t,t_0) X_j(t_0) X_k(t_0) \quad ; \quad \theta_{jk}(t,t_0) = cov\{R_j(t,t_0), R_k(t,t_0)\} \qquad (6.9)$$

One may consider Markowitz variance *Θ*$_M$*(t,t₀)* (6.9) as a zero approximation *χ=0* of the Taylor series of the market-based variance *Θ(t,t₀)* (5.10; 6.8). Indeed, if one assumes that all volumes $U_j(t_i)$ of the consecutive trades with all securities *j=1,..J* are constant, then the volumes $W(t_i)$ (3.4) of trades are also constant, and their coefficients of variation *χ* equal zero. The constant *a* in (6.8) describes the dependence (6.10) of the covariance *φ* (5.12) on the coefficients of variation *ψ* and *χ* of the values $Q(t_i)$ and volumes $W(t_i)$ of trades with the market portfolio:

$$\varphi = \frac{cov\{Q(t),W(t)\}}{Q(t;1)W(t;1)} = a \cdot \psi \cdot \chi \quad ; \quad -1 \leq a \leq 1 \qquad (6.10)$$

The relations (6.10) follow from the Cauchy-Schwarz-Bunyakovskii inequality (Shiryaev, 1999, p 123; Olkhov, 2025a; 2025b):

$$\frac{cov\{Q(t),W(t)\}}{cov^{\frac{1}{2}}\{Q(t),Q(t)\}cov^{\frac{1}{2}}\{W(t),W(t)\}} = a \quad ; \quad -1 \leq a \leq 1$$

$$\frac{cov\{Q(t),W(t)\}}{Q(t;1)W(t;1)} = a \frac{cov^{1/2}\{Q(t),Q(t)\}}{Q(t;1)} \frac{cov^{1/2}\{W(t),W(t)\}}{W(t;1)} = a \cdot \psi \cdot \chi$$

The Taylor series of the market-based variance *Θ*$_m$*(t,t₀)* (5.8) of all market trades has a form similar to (6.8):

$$\Theta_m(t,t_0) = \Theta_{Mm}(t,t_0) - 2 a_m \Theta_{Mm}^{1/2}(t,t_0) R_m(t,t_0) \chi_m + [R_m^2(t,t_0) - \Theta_{Mm}(t,t_0)] \chi_m^2 \qquad (6.11)$$

*R*$_m$*(t,t₀)* (5.7; 6.11) is the mean return of all market trades, and *Θ*$_{Mm}$*(t,t₀)* denotes Markowitz variance as a zero approximation for the case *χ*$_m$*=0*. The constant *a*$_m$ in (6.12) describes the dependence of covariance *φ*$_m$ (5.9) on the coefficients of variation *ψ*$_m$ and *χ*$_m$ of the values $C_m(t_i)$ and volumes $Um(t_i)$ of all market trades:

$$\varphi_m = a_m \cdot \psi_m \cdot \chi_m \quad ; \quad -1 \leq a_m \leq 1 \qquad (6.12)$$

In App. B we describe the difference (B.15) between Markowitz variance *Θ*$_M$*(t,t₀)* (6.8; 6.9) of the market portfolio and Markowitz variance *Θ*$_{Mm}$*(t,t₀)* (6.11) of all market trades.

$$\Theta_M(t,t_0) - \Theta_{Mm}(t,t_0) = \sum_{j,k=1}^{J} \theta_{jk}(t,t_0) \cdot X_j(t_0) \cdot X_k(t_0) \left[1 - \frac{x_j(t)}{x_j(t_0)} \frac{x_k(t)}{x_k(t_0)}\right] \qquad (6.13)$$

We recall that equation (6.13; B.15) describes the simplified model of real market trades when all volumes of consecutive trades with all market securities are assumed to be constant during the averaging interval $\Delta$ (2.10).

The distinctions (6.13) between Markowitz variances $\Theta_M(t,t_0)$ (6.8; 6.9) of the market portfolio and $\Theta_{Mm}(t,t_0)$ (6.11) of all market trades can be followed (App. C.) by the distinctions between the coefficient of variation $\chi$ (5.12) of the volumes $W(t_i)$ (3.4) of the consecutive trades with the market portfolio and the coefficient of variation $\chi_m$ (5.9) of the volumes $U_m(t_i)$ (4.1) of the consecutive trades with all market securities. The equation (C.9) presents the dependence of $\chi^2$ (5.12) on the coefficient of variation $\chi_m$ (5.9), and we refer to App. C for details:

$$1 + \chi^2 = (1 + \chi_m^2) \cdot (1 + \chi_\gamma^2) \cdot (1 + b\, \xi_\gamma\, \xi_m) \tag{6.14}$$

A rather complex dependence (6.14) reveals that $\chi^2$ (5.12) depends not only on the coefficient of variation $\chi_m$ (5.9) of the volumes $U_m(t_i)$ (4.1) of the consecutive trades with all market securities but also on the coefficients of variation $\xi_m$ (C.8) of squares of the volumes $U_m^2(t_i)$ (4.1). Moreover, the coefficient of variation $\chi$ (5.12) depends on the coefficient of variation $\chi_\gamma$ (C.4) of proportionality factor $\gamma(t_i)$ (C.1) that at time $t_i$ ties up the random volumes $W(t_i)$ (3.4) and $U_m(t_i)$ (4.1) and on the coefficient of variation $\xi_\gamma$ (C.8) of squares of $\gamma^2(t_i)$ (C.1).

In the next section we consider a few simple limiting cases that follow from (6.8; 611; 6.13; 6.14).

## 7.  Simple limiting cases

At first, let us consider the Taylor series (6.8; 6.11) for the simple case when the covariances $\varphi_m$ (5.9; 6.12) and $\varphi$ (5.12; 6.10) equal zero. Let us introduce the coefficients of variation $\mu(\psi,\chi,\varphi)$ (7.1) of the variances $\Theta(t,t_0)$ (5.11) of the market portfolio and $\mu_m(\psi_m,\chi_m,\varphi_m)$ (7.1) of variance $\Theta_m(t,t_0)$ (5.8) of all trades in the market:

$$\mu(\psi,\chi,\varphi) = \frac{\Theta(t,t_0)}{R^2(t,t_0)} = \frac{\psi^2 - 2\varphi + \chi^2}{1+\chi^2} \quad ; \quad \mu_m(\psi_m,\chi_m,\varphi_m) = \frac{\Theta_m(t,t_0)}{R_m^2(t,t_0)} = \frac{\psi_m^2 - 2\varphi_m + \chi_m^2}{1+\chi_m^2} \tag{7.1}$$

The similar case was described by Olkhov (2025b). If $\varphi=a=0$ (6.10), then from (B.3) the Taylor series (6.8) of the market portfolio takes the form:

$$\Theta(t,t_0) = \{\psi_0^2 + [1-\psi_0^2] \cdot \chi^2\} \cdot R^2(t,t_0) \quad ; \quad \mu(\psi,0,0) = \psi_0^2 = \frac{cov\{Q(t),Q(t)\}}{Q^2(t;1)}\Big|_{\chi^2=0} \tag{7.2}$$

The Taylor series (6.11) for the variance $\Theta_m(t,t_0)$ (5.8; 5.9) of all trades in the market gives:

$$\Theta_m(t,t_0) = \{\psi_{0m}^2 + [1-\psi_{0m}^2] \cdot \chi_m^2\} \cdot R_m^2(t,t_0) \quad ; \quad \mu_m(\psi_m,0,0) = \psi_{0m}^2 = \frac{cov\{C_m(t),C_m(t)\}}{C_m^2(t;1)}\Big|_{\chi_m^2=0} \tag{7.3}$$

If the coefficient of variation $\mu(\psi,0,0)=\psi_0^2<<\chi^2<1$ of the market portfolio is much less than a unit, then $1-\psi_0^2\sim1$ and the variance $\Theta(t,t_0)$ (7.2) of the market portfolio takes the form:

$$\Theta(t,t_0) = (\psi_0^2 + \chi^2) \cdot R^2(t,t_0) = \Theta_M(t,t_0) + \chi^2 \cdot R^2(t,t_0) \tag{7.4}$$

The condition $\mu(\psi,0,0)=\psi_0^2<<1$ means that Markowitz variance $\Theta_M(t,t_0)$ (6.9) of the market portfolio is very low, and the investors may imagine that their market portfolios have low risks. However, the growth of the coefficient of variation $\chi^2$ so that $\mu(\psi,0,0)=\psi_0^2<<\chi^2$ may significantly increase the portfolio variance $\Theta(t,t_0)$ (7.4) compared with Markowitz's estimate of the variance $\Theta_M(t,t_0)$ (6.9). The same considerations for (7.3) give:

$$\Theta_m(t,t_0) = \Theta_{Mm}(t,t_0) + \chi_m^2 \cdot R_m^2(t,t_0) \tag{7.5}$$

At first glance it seems that the high variance $\Theta_m(t,t_0)$ (7.5) of trades with all securities in the market should be accompanied by the high variance $\Theta(t,t_0)$ (7.4) of the market portfolio. As we show in App. D, is not always so. In App. D, we consider very simple toy models to illustrate the possible distinctions between the variances $\Theta_m(t,t_0)$ (7.5) and $\Theta(t,t_0)$ (7.4). We regard trivial models of the stock market that consist of only two securities and assume that the volumes of consecutive trades with securities are constant. These toy models reveal that the different values (D.5; D.11) of the relative numbers $x_j(t_0)$ (2.3) of shares in the market portfolio, the relative numbers $x_j(t)$ (4.4) of the volumes $U_j$ of trades with securities (D.5; D.11), and the relative amounts $X_j(t_0)$ (2.4) invested into securities (D.6; D.12) may cause that Markowitz variance $\Theta(t,t_0)$ (7.4) of the market portfolio may occur several times less (D.10), or several times more (D.14) than Markowitz variance $\Theta_{Mm}(t,t_0)$ (7.5) of trades with all securities in the market. We recall that Markowitz's (1952) variance (6.9) ignores the impact of the random variances of the volumes of consecutive trades. In App. D. we briefly study a case where all volumes $U_m(t_i)$ (4.1; D.17) of trades with all securities in the market are constant, but the volumes $W(t_i)$ (3.4; D.18; D.19) of trades with the market portfolio are random. In this case the coefficient of variation $\chi^2$ (5.12) of the volumes $W(t_i)$ may additionally increase the difference between the low variance $\Theta_{Mm}(t,t_0)$ (D.13) and risks of all trades in the market and the relatively high variance $\Theta(t,t_0)$ (D.14; D.22) and risks of the market portfolio.

The investors and portfolio managers should consider the mutual relations between the variances and risks of their market portfolios and the current performance of the trades with all market securities during the averaging interval $\Delta$ (2.10). The forecasts of these relations are necessary but extremely challenging.

## 8. The duration of the averaging interval and the forecasting problems

### 8.1 *The duration of the averaging interval*

The duration of the averaging interval $\varDelta$ (2.10) plays the crucial role in the assessments of the variance and risks of the market portfolio and of the trades with all securities in the market. To derive a reliable statistical assessment of the mean return and the variance of his portfolio, the investor should choose the duration of the averaging interval $\varDelta$ during which the total volumes $U_{\Sigma j}(t)$ (2.13) of market trades with each security $j=1,...J$ of his portfolio should be much more (8.1) than the number of shares $W_j(t_0)$ of these securities in his portfolio:

$$U_{\Sigma j}(t) = \sum_{i=1}^{N} U_j(t_i) \gg W_j(t_0) \tag{8.1}$$

The choice of the duration of $\varDelta$ (2.10) and (8.1) should guarantee that if the investor decides to sell his portfolio, his sale should not disturb the assessments of the mean return and the variance of his portfolio that the investor makes while he observes the current trades with his securities in the market. Hence, the number of shares of securities $j=1,...J$ in the portfolio should be small compared with the total volume of trades with security $j$ during $\varDelta$. We propose that the number of shares of each security $j=1,...J$ should be less than 3-5% of the total volume $U_{\Sigma j}(t)$ (2.13) of trades. If so, the investor may hope that the sale of his shares during the interval of the same duration as $\varDelta$ (2.10) will not significantly penetrate the market. That condition highlights that market portfolios with the different initial value $Q_\Sigma(t_0)$ (2.6) and hence the different number of shares $W_j(t_0)$ (2.4; 2.5) should have different durations of the averaging intervals $\varDelta$ (2.10). Simply speaking, to assess the average return and the variance of the market portfolio with the initial value $Q_\Sigma(t_0)$~1K\$ and with the value $Q_\Sigma(t_0)$~100M\$, one should choose the averaging intervals with different durations. Otherwise, the investor's assessments of the current returns and the variance of his portfolio and of the trades with the total volume $U_{\Sigma j}(t)$ (2.13) during the interval $\varDelta$ that are the same or less than the number of shares $W_j(t_0)$ of security $j$ will be a waste of time. The attempt to sell $W_j(t_0)$ shares during the interval with such duration will so much perturbate the market that all the estimates and the forecasts made on the basis of the observation of the trades will be completely wrong and misleading, and the investor may lose a lot of money.

That creates complex problems for market-based assessments of the market portfolios, which grow up with the increase of total initial value $Q_\Sigma(t_0)$. Indeed, the theoretical attempts to estimate market-based average return and the variance of the entire stock market are illusory. Our doubts are based on the collision between the conventional, "regulatory" assessments of the capitalization of the company *j* with shares outstanding $W_j$ and the market-based assessment of the portfolio with the total initial value $Q_\Sigma(t_0)$. As an example, let us consider Apple Inc. and roughly assess its market capitalization. According to usual "regulatory" rules, the capitalization of the company with $W_j$ shares outstanding is determined on a daily basis by the closing prices, which are calculated as VWAP during the last 30 minutes of the trading hours. NASDAQ normal trading hours are 9:30 am to 4 pm. Thus, the closing price is calculated during the period that equals 1/13 of normal trading hours. For Apple Inc., the share turnover ratio varies around 0,8-1% of shares outstanding $W_a$ during the trading hours. The capitalization of Apple Inc. $CapAAPL = p_a \cdot W_a$ is determined by the closing price $p_a$ that is calculated as VWAP during 1/13 of trading hours on the basis of time series with trade volume that represents less than 0,1% of outstanding shares $W_a$.

Simple estimates show that at least 100 trading days, or half of a year's working days, are required to get the turnover equal to the total shares outstanding $W_a$. Hence, the imaginary sale of the portfolio of all Apple Inc. shares outstanding in pieces of 3-5% in the "business-as-usual" market may require more than 10 years. It is obvious that such a case has nothing in common with real investments. This problem is solved by the OTC trading of huge equity stakes of 10-30% or more, under the price agreed upon by the parties involved in the deal. However, the OTC deals considerably rely upon the current assessments of the returns and the variance on the basis of the market trades in the exchange, such as NASDAQ, NYSE, etc. The description of the mutual dependence of the Exchange and OTC trading requires further studies.

**8.2** *The predicting of the mean return and variances*

Numerous researchers study the problems of forecasting the mean returns and the variances. Most studies are based on the consideration of their time series (Lakonishok, 1980; Nelson and Kim, 1990; Kandel and Stambaugh, 1995; Campbell and Yogo, 2003; Golez and Koudijs, 2017; Kelly, Malamud, and Zhou, 2022). As well as Markowitz's (1952) variance, all these studies and forecasting of the time series of returns and variances implicitly and latently use the simplified market approximation and unintentionally propose that all volumes of the consecutive trades during the averaging interval are constant. However, the real markets reveal high irregular or random variations of the volumes of consecutive trades. The excess simplification of constant trade volumes is one of the reasons for the low accuracy of predictions of return and variance that are based on their time series.

Sections 2-7 of our paper reveal that the returns and the variances of the current trades (5.7; 5.8) and of the market portfolio (5.10; 5.11) are determined by the time series of the values $C_j(t_i)$ and the volumes $U_j(t_i)$ (2.11-2.14) of consecutive trades with market securities $j=1,...J$. The origin of the complexity of the forecasting of returns and variances at a particular time horizon $T$ is explained by the complexity of the predictions of the time series of the values $C_j(t_i)$ and the volumes $U_j(t_i)$ (2.11-2.14) of market trades with market securities $j=1,...J$ at this horizon $T$ during the selected averaging interval.

Our description of the mean returns and the variances makes evident that one can substitute the high-frequency time series of consecutive trades (2.11-2.14) that occur with short time span $\varepsilon$ (2.10) with the aggregated time series of consecutive trades with the time span $\zeta$ of any required duration. One can consider the duration $\zeta$ between market trades that can be equal to $\zeta(1)=1$ min, $\zeta(2)=1$ hour, $\zeta(3)=1$ day, or whatever. To do that, one should sum the initial market time series of the values $C_j(t_i)$ and the volumes $U_j(t_i)$ (2.11-2.14) of consecutive trades with securities $j=1,...J$ during the consecutive time spans $\tau_k$ of selected duration $\zeta$:

$$C_j(\tau_k) = \sum_{i=1}^{N} C_j(t_{k,i}) \ ; \ U_j(\tau_k) = \sum_{i=1}^{N} U_j(t_{k,i}) \ ; \ C_j(\tau_k) = p_j(\tau_k) U_j(\tau_k) \qquad (8.2)$$

$$t_{k,i} \in \tau_k \ , \ i = 1,..N \ , \ k = 1,...K \qquad (8.3)$$

In (8.3) $t_{k,i}$ denotes the time inside the interval $\tau_k$. One can reproduce all results of our paper for the averaging interval of duration $\Delta_\tau$:

$$\Delta_\tau = \left[t - \frac{\Delta_\tau}{2}; t + \frac{\Delta_\tau}{2}\right] \ ; \ \tau_k \in \Delta_\tau \ ; \ k = 1,...K \ ; \ K \cdot \zeta = \Delta_\tau \qquad (8.4)$$

The averaging of the time series (8.2) during $\Delta_\tau$ (8.4) gives consecutive approximations of the mean returns and the variances of the market portfolio and the trades with all market securities for different time spans $\zeta$ and different averaging intervals $\Delta_\tau$. To forecast the mean return and the variance of the portfolio at the time horizon $T$ in the model with selected time span $\zeta$ and the averaging interval $\Delta_\tau$, the investors or portfolio managers should predict the time series of the values $C_j(\tau_k)$ and the volumes $U_j(\tau_k)$ of consecutive market trades. Such forecasts are almost equal to the predictions of macroeconomic variables that impact and that depend upon the business activity and business performance of all $j=1,...J$ companies and corporations whose stocks are traded in the market exchange with the values $C_j(\tau_k)$ and the volumes $U_j(\tau_k)$. The forecast of the returns and the variances of the market portfolios depend upon the forecasts of macroeconomic variables. The forecasts of the returns and the variances of the portfolios that are composed of few securities $j=1,...J_p$, where $J_p<<J$ is much less than the number $J$ of all securities in the market, require the prediction of the macroeconomic environment that determines the economic activity of $j=1,...J_p$ companies and governs the random time series of the values $C_j(\tau_k)$ and the volumes $U_j(\tau_k)$ of trades in the market. The prices, returns, and variances of market stocks are the consequences of the business performance of the corresponding companies. The predictions of random market trade time series that depend on corresponding macroeconomic forecasts determine the main problem, the major obstacle for the reliable forecasts of the returns and variances. These critical essential problems may be worthwhile for BlackRock's Aladdin, which should be complemented by the much more powerful *"Divine Augur"* to be able to perform complex probabilistic and dynamic prophecies of macroeconomic variables.

Meanwhile, the random variations of the values and the volumes of trades in commodities, retail, financial, energy, etc. markets determine the low bounds of uncertainty of macroeconomic observations (Olkhov, 2021; 2023b; 2024) and impact the accuracy of macroeconomic forecasts. Macroeconomic complexity limits the accuracy of the forecasts (Olkhov, 2023b; 2023c; 2024) of the probability of prices and returns at best by the Gaussian distributions. But even the achieving of that accuracy may not be any time soon and require a lot of further studies and efforts.

Future returns and variances are safely hidden from investors.

## 9. Conclusion

This paper gives the novel unified description of the returns and variances of the trades with a particular security $j=1,J$; the trades with all securities in the market as the trades with a single security; and the trades with the market portfolio as trades with a single security.

There is no difference between the description of the returns and variances of the trades with a single security $j$, of the trades with all securities in the market, and of the market portfolio.

One can use the same methods to describe the returns and variances of any portfolio that the investor collected of $j=1,...J_p$ securities, where $J_p<J$ is less than the number $J$ of all securities in the market. The same methods describe the returns and variances of the trades with securities of any portfolio.

The investors and portfolio managers should account for the impact of the duration of the averaging interval on the reliability and confidence of their estimates of the portfolio return and variance. Too short averaging intervals may give the assessments that will not occur while the investor decides to sell his portfolio.

The predictions of the returns and variances of the market portfolios depend on the duration of the averaging interval and on the macroeconomic forecasting. The random variations of the values and the volumes of the trades in commodities, retail, financial, energy markets, etc., determine the uncertainty of macroeconomic observations, which limits the accuracy of macroeconomic forecasts at best by the Gaussian distributions. That establishes the upper limits on the accuracy of the predictions of the returns and variances of a single security, of portfolios of $J_p$ securities, and of market portfolios. The predictions of the returns and variances as a part of macroeconomic forecasting are the problems for such market majors as BlackRock, JP Morgan, and the U.S. Fed. However, even BlackRock's Aladdin should be complemented by the much more powerful *"Divine Augur,"* which will perform complex probabilistic and dynamic prophecies of macroeconomic variables. Those are the far future goals.

## Appendix A. Market-based variance

One can determine probability by the set of consistent statistical moments (Shiryaev, 1999; Shreve, 2004). The market-based VWAP price *s(t;1)* (3.6) determines the *1$^{st}$* price statistical moment. Olkhov (2022; 2023a; 2025a) showed that the market-based variance of price *Φ(t)*, or the *2$^{nd}$* central price statistical moment that is consistent with VWAP price *s(t;1)* (3.6) during *Δ,* equals:

$$\Phi(t) = \frac{1}{W_\Sigma(t;2)} \sum_{i=1}^{N} \big(s(t_i) - s(t;1)\big)^2 W^2(t_i) \quad ; \quad W_\Sigma(t;2) = \sum_{i=1}^{N} W^2(t_i) \qquad (A.1)$$

One can present the variance of price *Φ(t)* (A.1) as follows:

$$\Phi(t) = \Phi(t;2) - 2\Phi(t;1)s(t;1) + s^2(t;1) \qquad (A.2)$$

The functions *Φ(t;2)* and *Φ(t;1)* take the forms:

$$\Phi(t;2) = \frac{1}{W_\Sigma(t;2)} \sum_{i=1}^{N} s^2(t_i) W^2(t_i) = \frac{1}{W(t;2)} \frac{1}{N} \sum_{i=1}^{N} Q^2(t_i) = \frac{Q(t;2)}{W(t;2)} \qquad (A.3)$$

$$\Phi(t;1) = \frac{1}{W_\Sigma(t;2)} \sum_{i=1}^{N} s(t_i) W^2(t_i) = \frac{1}{W(t;2)} \frac{1}{N} \sum_{i=1}^{N} Q(t_i) W(t_i) = \frac{E[Q(t_i)\,W(t_i)]}{W(t;2)} \qquad (A.4)$$

In (A.3) we use (5.6; 5.12) and define average squares of the values *Q(t;2)* and volumes *W(t;2)*:

$$Q(t;2) = \frac{1}{N} \sum_{i=1}^{N} Q^2(t_i) = Q^2(t;1)[1 + \psi^2]; \quad W(t;2) = \frac{W_\Sigma(t;2)}{N} = W^2(t;1)[1 + \chi^2] \qquad (A.5)$$

The joint average *E[Q(t$_i$),U(t$_i$)]* of values and volumes and (5.5; 5.12) give:

$$E[Q(t_i)\,W(t_i)] = Q(t;1)W(t;1)[1 + \varphi] \qquad (A.6)$$

The substitution of (A.3-A.6) into (A.2), gives:

$$\Phi(t) = \frac{Q^2(t;1)[1+\psi^2] - 2Q(t;1)W(t;1)[1+\varphi]s(t)}{W^2(t;1)[1+\chi^2]} + \frac{Q^2(t;1)}{W^2(t;1)}$$

Simple transformations and use of (3.6; 3.7; 5.12) give the variance of price $\Phi(t)$ (A.1):

$$\Phi(t) = \frac{\psi^2 - 2\varphi + \chi^2}{1+\chi^2} s^2(t;1) \tag{A.7}$$

From (A.7), obtain the market-based variance $\Theta(t,t_0)$ of return:

$$\Theta(t,t_0) = \frac{\Phi(t)}{s^2(t_0)} = \frac{\psi^2 - 2\varphi + \chi^2}{1+\chi^2} R^2(t,t_0) \tag{A.8}$$

If all trade volumes $W(t_i)$ during $\Delta$ are constant, then $\chi^2 = \varphi = 0$ and the variance $\Theta(t,t_0)$ equals:

$$\Theta(t,t_0) = \psi^2 \cdot R^2(t,t_0) = \frac{1}{N}\sum_{i=1}^{N}[R(t_i,t_0) - R(t,t_0)]^2 \tag{A.9}$$

One can find the additional justifications in (Olkhov, 2022; 2023a; 2025a).

## Appendix B.   Markowitz's variances of portfolio and of market trades

In this App. we consider Markowitz's expressions of the variance for the market portfolio and of all market trades in the approximation where all trade volumes $U_j(t_i)=U_j$ of trades with securities $j=1,...J$ are assumed constant. Hence, the volumes $W(t_i)=W$ of trades with the portfolio and the volumes $U_m(t_i)=U_m$ of all market trades also are constant. Thus, Markowitz's expressions of the variance ignore the effects of fluctuations of the consecutive trade volumes. The total volumes of trades with securities, with the entire market, and with the market portfolio during $\Delta$ obey relations:

$$U_{\Sigma j}(t) = N \cdot U_j \; ; \; U_{\Sigma m}(t) = N \cdot U_m \; ; \; u_j = \frac{W_j(t_0)}{U_{\Sigma j}(t)} U_j = \frac{W_j(t_0)}{N} \; ; \; W_\Sigma(t_0) = N \cdot W \tag{B.1}$$

From (2.14; 3.6; 4.3) and (B.1), obtain the expressions of mean prices in the assumption that all volumes $U_j(t_i)=U_j$ of trades are constant:

$$p_j(t;1) = \frac{1}{N}\sum_{i=1}^{N} p_j(t_i) \; ; \; s(t;1) = \frac{1}{N}\sum_{i=1}^{N} s(t_i) \; ; \; s_m(t;1) = \frac{1}{N}\sum_{i=1}^{N} s_m(t_i) \tag{B.2}$$

If the trade volumes $W(t_i)=W$ and $U_m(t_i)=U_m$ are constant, then their coefficients of variation $\chi=0$ and the covariance $\varphi=0$ (5.12) and $\chi_m=0$ and the covariance $\varphi_m=0$ (5.9) equal zero. If so, the variance of the market portfolio $\Theta(t,t_0)$ (5.11; 6.8) and the variance $\Theta_m(t,t_0)$ (5.8; 6.11) of all market trades take the form:

$$\Theta(t,t_0) = \Theta_M(t,t_0) = \psi_0^2 \cdot R^2(t,t_0) \quad ; \quad \Theta_m(t,t_0) = \Theta_{Mm}(t,t_0) = \psi_{0m}^2 \cdot R_m^2(t,t_0) \tag{B.3}$$

$$\psi_0^2 = \frac{cov\{Q(t),Q(t)\}}{Q^2(t;1)}\Big|_{\chi^2=0} \quad ; \quad \psi_{0m}^2 = \frac{cov\{C_m(t),C_m(t)\}}{C_m^2(t;1)}\Big|_{\chi_m^2=0}$$

From (5.12), (3.4), (3.1), and (B.1), obtain:

$$Q(t;1) = \frac{1}{N}\sum_{i=1}^{N} s(t_i)\,W = s(t;1)\cdot W \quad ; \quad s(t;1) = \frac{1}{N}\sum_{i=1}^{N} s(t_i) \tag{B.4}$$

We highlight that in the approximation where all trade volumes $W$ are constant, VWAP $s(t;1)$ (2.14) takes the form (B.4). The mean prices $p_j(t;1)$ (2.14) and $s_m(t;1)$ (4.3) take the form similar to (B.4). To assess $\psi^2$ (5.12), let us use (3.1; 3.4; B.1) and transform the covariance:

$$cov\{Q(t), Q(t)\} = \frac{1}{N}\sum_{i=1}^{N}\sum_{j,k=1}^{J}\left(c_j(t_i) - c_j(t;1)\right)\left(c_k(t_i) - c_k(t;1)\right) = \quad (B.5)$$

$$= \sum_{j,k=1}^{J}\frac{W_j(t_0)}{N}\frac{W_k(t_0)}{N}\frac{1}{N}\sum_{i=1}^{N}\left(p_j(t_i) - p_j(t;1)\right)\left(p_k(t_i) - p_k(t;1)\right) = \sum_{j,k=1}^{J}\frac{W_j(t_0)}{N}\frac{W_k(t_0)}{N}\sigma_{jk}^2(p)$$

In (B.5), $\sigma_{jk}(p)$ denotes the covariance of prices $p_j(t_i)$ and $p_k(t_i)$ during $\Delta$:

$$\sigma_{jk}^2(p) = \frac{1}{N}\sum_{i=1}^{N}\left(p_j(t_i) - p_j(t;1)\right)\left(p_k(t_i) - p_k(t;1)\right) \quad ; \quad p_j(t;1) = \frac{1}{N}\sum_{i=1}^{N}p_j(t_i) \quad (B.6)$$

From (5.12; B.1; B.4-B.6), and (6.2), obtain:

$$\psi_0^2 = \frac{1}{s^2(t;1)\cdot W^2}\sum_{j,k=1}^{J}\frac{W_j(t_0)}{N}\frac{W_k(t_0)}{N}\sigma_{jk}^2(p) = \frac{1}{s^2(t;1)}\sum_{j,k=1}^{J}\sigma_{jk}^2(p)\cdot x_j(t_0)\cdot x_k(t_0) \quad (B.7)$$

From (5.10) and (6.5), obtain:

$$\Theta_M(t, t_0) = \frac{1}{s^2(t;1)}\sum_{j,k=1}^{J}\sigma_{jk}^2(p)\cdot x_j(t_0)\cdot x_k(t_0)\frac{s^2(t;1)}{s^2(t_0)}$$

$$\Theta_M(t, t_0) = \sum_{j,k=1}^{J}\frac{\sigma_{jk}^2(p)}{p_j(t_0)p_k(t_0)}\cdot\frac{p_j(t_0)}{s(t_0)}x_j(t_0)\cdot\frac{p_k(t_0)}{s(t_0)}x_k(t_0) = \sum_{j,k=1}^{J}\theta_{jk}(t,t_0)\cdot X_j(t_0)X_j(t_0) \quad (B.8)$$

In (B.8) functions $\theta_{jk}(t,t_0)$ denote the covariances between returns (5.1) of securities $j$ and $k$:

$$\theta_{jk}(t, t_0) = \frac{1}{N}\sum_{i=1}^{N}\frac{\left(p_j(t_i) - p_j(t;1)\right)}{p_j(t_0)}\frac{\left(p_k(t_i) - p_k(t;1)\right)}{p_k(t_0)} = \frac{1}{N}\sum_{i=1}^{N}[R_j(t_i,t_0) - R_j(t,t_0)]\cdot[R_j(t_i,t_0) - R_j(t,t_0)] \quad (B.9)$$

We highlight that the above expressions coincide with the definitions of portfolio variance $\Theta_M(t,t_0)$ (B.8) and the covariances (B.9) between returns of securities $j$ and $k$, and functions $X_j(t_0)$ (2.3; 4.4; 6.5) have the meaning of the relative amount invested into security $j$.

The above results reveal that Markowitz's expression of the portfolio variance $\Theta_M(t,t_0)$ (B.8) describes the simplified model of the real markets when the volumes of consecutive trades with all market securities are assumed to be constant during the averaging interval.

Now we consider the variance $\Theta_{Mm}(t,t_0)$ (B.3) of all market trades during $\Delta$ (2.10) and show that its expression is different from (B.8). From (4.1; B.1), obtain mean values $C_m(t;1)$:

$$C_m(t;1) = \frac{1}{N}\sum_{i=1}^{N}s_m(t_i)U_m = s_m(t;1)\cdot U_m \quad ; \quad s_m(t;1) = \frac{1}{N}\sum_{i=1}^{N}s_m(t_i) \quad (B.10)$$

To assess $\psi_{0m}^2$ (5.9), we use (2.11; 4.1; B.1) and (B.6), and transform the covariance:

$$cov\{C_m(t), C_m(t)\} = \frac{1}{N}\sum_{i=1}^{N}\sum_{j,k=1}^{J}\left(C_j(t_i) - C_j(t;1)\right)\left(C_k(t_i) - C_k(t;1)\right) = \quad (B.11)$$

$$= \sum_{j,k=1}^{J}U_jU_k\frac{1}{N}\sum_{i=1}^{N}\left(p_j(t_i) - p_j(t;1)\right)\left(p_k(t_i) - p_k(t;1)\right) = \sum_{j,k=1}^{J}U_jU_k\sigma_{jk}^2(p)$$

From (5.12; B.1; B.4-B.6), and (6.2), obtain:

$$\psi_{0m}^2 = \frac{1}{s_m^2(t;1)\cdot U_m^2} \sum_{j,k=1}^{J} U_j U_k \, \sigma_{jk}^2(p) = \frac{1}{s_m^2(t;1)} \sum_{j,k=1}^{J} \sigma_{jk}^2(p) \cdot x_j(t) \cdot x_k(t) \quad (B.12)$$

From (B.2), obtain that the relative numbers $x_j(t)$ (4.4) of the trade volumes $U_{\Sigma j}(t)$ with security $j$ with respect to the total trade volume $U_{\Sigma m}(t)$ during $\Delta$ equal the ratio (B.13) of constant trade volume $U_j$ to the constant trade volume $U_m$. The ratio $X_j(t)$ (4.4) of the total values $C_{\Sigma j}(t)$ (2.13) of trade with security $j$ to the total value $C_{\Sigma m}(t)$ (4.3) of all market trades takes the form:

$$x_j(t) = \frac{U_{\Sigma j}(t)}{U_{\Sigma m}(t)} = \frac{U_j}{U_m} \; ; \quad X_j(t) = \frac{C_{\Sigma j}(t)}{C_{\Sigma m}(t)} = \frac{p_j(t;1)\cdot U_j}{s_m(t;1)\cdot U_m} = \frac{p_j(t;1)}{s_m(t;1)} \cdot x_j(t) \quad (B.13)$$

Finaly, obtain:

$$\Theta_{Mm}(t,t_0) = \psi_{0m}^2 \cdot R_m^2(t,t_0) = \frac{1}{s^2(t_0)} \sum_{j,k=1}^{J} \sigma_{jk}^2(p) \cdot x_j(t) \cdot x_k(t)$$

$$\Theta_{Mm}(t,t_0) = \sum_{j,k=1}^{J} \theta_{jk}(t,t_0) \cdot \frac{p_j(t_0)}{s(t_0)} x_j(t) \cdot \frac{p_k(t_0)}{s(t_0)} x_k(t) \quad (B.14)$$

The difference between Markowitz's expressions of the variance $\Theta_M(t,t_0)$ (B.8) of the market portfolio and the variance $\Theta_{Mm}(t,t_0)$ (B.14), takes the form (B.15):

$$\Theta_M(t,t_0) - \Theta_{Mm}(t,t_0) = \sum_{j,k=1}^{J} \theta_{jk}(t,t_0) \cdot X_j(t_0) \cdot X_k(t_0) \left[1 - \frac{x_j(t)}{x_j(t_0)} \frac{x_k(t)}{x_k(t_0)}\right] \quad (B.15)$$

## Appendix C. Coefficients of variation of the volumes of trades

Here we describe the dependence of the coefficient of variation $\chi$ (5.12) of random volumes $W(t_i)$ (3.4) of the consecutive trades with the market portfolio on the coefficient of variation $\chi_m$ (5.9) of random volumes $U_m(t_i)$ (4.1) of trades with all market securities.

Let us define the coefficient $\gamma(t_i)$ that describe the dependence of the volume $W(t_i)$ (3.4) of trade with the market portfolio at time $t_i$ on the volume $U_m(t_i)$ (4.1) of trade with all securities:

$$W(t_i) = \gamma(t_i) \cdot U_m(t_i) \; ; \; W(t;1) = \gamma(t;1)\, U_m(t;1) \; ; \; \gamma(t;1) = \frac{W(t;1)}{U_m(t;1)} = \frac{1}{U_{\Sigma m}(t)} \sum_{i=1}^{N} \gamma(t_i) \cdot U_m(t_i) \quad (C.1)$$

We recall that functions $W(t;1)$, $U_m(t;1)$, and $\gamma(t;1)$ in (C.1) denote the mean values of random volumes $W(t_i)$ (3.4), $U_m(t_i)$ (4.1), and random coefficient $\gamma(t_i)$ during the averaging interval $\Delta$ (2.10). Random coefficients $\gamma(t_i)$ tie up the random volumes $W(t_i)$ (3.4) of consecutive trades with the market portfolio of the investor and random volumes $U_m(t_i)$ (4.1) of trades with all securities. From (5.12), obtain:

$$\chi^2 = \frac{cov\{W(t),W(t)\}}{W^2(t;1)} = \frac{E[W^2(t_i)] - W^2(t;1)}{W^2(t;1)} \rightarrow 1 + \chi^2 = \frac{E[W^2(t_i)]}{W^2(t;1)} \quad (C.2)$$

$$E[W^2(t_i)] = E[\gamma^2(t_i) \cdot U_m^2(t_i)] = E[\gamma^2(t_i)] \cdot E[U_m^2(t_i)] + cov\{\gamma^2(t_i), U_m^2(t_i)\} \quad (C.3)$$

Let us define the coefficient of variation $\chi_\gamma$ of $\gamma(t_i)$ alike to the coefficient of variation $\chi$ (5.12):

$$\chi_\gamma^2 = \frac{cov\{\gamma(t),\gamma(t)\}}{\gamma^2(t;1)} = \frac{E[\gamma^2(t_i)]}{\gamma^2(t;1)} - 1 \quad ; \quad \gamma(t;2) = E[\gamma^2(t_i)] \tag{C.4}$$

The function $\gamma(t;2)$ (C.4) defines the mean squares of the coefficient $\gamma(t_i)$. From (5.9; C.1; C.3; C.4), obtain:

$$1 + \chi^2 = \frac{[\gamma^2(t_i)]}{\gamma^2(t;1)} \cdot \frac{E[U_m^2(t_i)]}{U_m^2(t;1)} + \frac{cov\{\gamma^2(t_i),U_m^2(t_i)\}}{\gamma^2(t;1)U_m^2(t;1)}$$

$$1 + \chi^2 = (1 + \chi_m^2)(1 + \chi_\gamma^2) + \frac{cov\{\gamma^2(t_i),U_m^2(t_i)\}}{\gamma^2(t;1)U_m^2(t;1)} \tag{C.5}$$

Let us transform the covariance in (C.5) of squares of $\gamma^2(t_i)$ and $U_m^2(t_i)$ as follows:

$$\frac{cov\{\gamma^2(t_i),U_m^2(t_i)\}}{\gamma^2(t;1)U_m^2(t;1)} = \frac{\gamma(t;2)}{\gamma^2(t;1)} \cdot \frac{U_m(t;2)}{U_m^2(t;1)} \cdot \frac{cov\{\gamma^2(t_i),U_m^2(t_i)\}}{\gamma(t;2)U_m(t;2)} = (1 + \chi_m^2)(1 + \chi_\gamma^2)\frac{cov\{\gamma^2(t_i),U_m^2(t_i)\}}{\gamma(t;2)U_m(t;2)} \tag{C.6}$$

We define $\Omega$ (C.7) the covariance between squares of $\gamma^2(t_i)$ and $U_m^2(t_i)$ and use the Cauchy-Schwarz-Bunyakovskii inequality (Shiryaev, 1999, p 123; Olkhov, 2025a; 2025b) to present it as:

$$\Omega = \frac{cov\{\gamma^2(t_i),U_m^2(t_i)\}}{\gamma(t;2) \cdot U_m(t;2)} = \omega \cdot \xi_\gamma \cdot \xi_m \quad ; \quad -1 \leq \omega \leq 1 \tag{C.7}$$

$$1 + \xi_m^2 = \frac{E[U_m^4(t)]}{U_m^2(t;2)} \quad ; \quad 1 + \xi_\gamma^2 = \frac{E[\gamma^4(t)]}{\gamma^2(t;2)} \tag{C.8}$$

Functions $\xi_m$ and $\xi_\gamma$ in (C7; C.8) denote the coefficients of variation of squares $U_m^2(t_i)$ and $\gamma^2(t_i)$. From (C.5-C.8), obtain:

$$1 + \chi^2 = (1 + \chi_m^2) \cdot (1 + \chi_\gamma^2) \cdot (1 + \omega \cdot \xi_\gamma \cdot \xi_m) \tag{C.9}$$

The equation (C.9) reveals that the coefficient of variation $\chi$ (5.12) of random volumes $W(t_i)$ (3.4) of the consecutive trades with the market portfolio depends upon the coefficient of variation of random volumes $U_m(t_i)$ (4.1) of trade with all market securities and on the coefficient of variation of random coefficients $\gamma(t_i)$ (C.1). Moreover, $\chi$ (5.12) depends upon the coefficients of variation $\xi_m$ (C.8) of squares of $U_m^2(t_i)$ (4.1) and $\xi_\gamma$ (C.8) of squares of $\gamma^2(t_i)$ (C.1).

## Appendix D. A toy model of market portfolio and market trades

Let us consider toy models that highlight the possible distinctions between the variance of the market portfolio and the variance of the trades with all market securities during the averaging interval $\Delta$ (2.10) as illustrations of the possible problems with the assessments of the variances. We simplify the models as much as possible and consider the market that consists of two securities, *1* and *2*. We assume that the values, the volumes, and the prices of consecutive trades and the returns of securities *1* and *2* have zero covariances.

At first we consider the trivial model where all volumes $U_j(t_i)$ of consecutive trades with securities $j=1,2$ are constant. Even such a trivial model hides some nontrivial relations. We assume:

$$U_{\Sigma 1}(t) = N \cdot U_1 \ ; \ U_{\Sigma 2}(t) = N \cdot U_2 \ ; \ U_{\Sigma m}(t) = N \cdot (U_1 + U_2) \ ; \ U_1, U_2 - const \quad (D.1)$$

From (D.1), obtain that the volumes $U_m(t_i)$ (4.1) of trades with all securities and the volumes $W(t_i)$ (3.3; 3.4) of trades with the market portfolio also are constant:

$$U_m(t_i) = U_m = U_1 + U_2 \ ; \ W(t_i) = W = \frac{W_1(t_0)}{U_{\Sigma 1}(t)} U_1 + \frac{W_2(t_0)}{U_{\Sigma 2}(t)} U_2 = \frac{1}{N}(W_1(t_0) + W_2(t_0)) \quad (D.2)$$

From (D.2) follows that the coefficients of variation $\chi_m^2$ (5.9) and $\xi_m^2$ (C.8) of the volumes $U_m$ and the coefficients of variation $\chi^2$ (5.12) equal zero. Hence, the variance $\Theta_m(t,t_0)$ (5.8; 6.11; 74) takes the form of Markowitz variance $\Theta_{Mm}(t,t_0)$ (B.14):

$$\Theta_m(t, t_0) = \Theta_{Mm}(t, t_0) = \psi_{0m}^2 R_m^2(t, t_0) \ ; \quad \chi_m^2 = 0 \quad (D.3)$$

The variance $\Theta(t,t_0)$ (5.11) of the market portfolio equals to Markowitz variance $\Theta_M(t,t_0)$ (6.9):

$$\Theta(t, t_0) = \Theta_M(t, t_0) = \sum_{j,k=1}^{J} \theta_{jk}(t, t_0) X_j(t_0) X_k(t_0) = \psi_0^2 R^2(t, t_0) \ ; \ \chi^2 = 0 \quad (D.4)$$

To estimate the possible difference (B.15) between Markowitz variance $\Theta_M(t,t_0)$ (D.4) of the market portfolio and Markowitz variance $\Theta_{Mm}(t,t_0)$ (D.3) of the trades, let us consider simple numeric examples. Let us assume that the relative number of shares $x_j(t_0)$ (2.3) in the market portfolio and the relative number $x_j(t)$ (4.4) of the volumes $U_j$ of trades with security $j$ at time $t$ take the values (D.5):

$$x_1(t_0) = \frac{W_1(t_0)}{W_\Sigma(t_0)} = \frac{2}{3} \ ; \ x_2(t_0) = \frac{1}{3} \ ; \quad x_1(t) = \frac{U_1}{U_m} = \frac{1}{3} \ ; \ x_2(t) = \frac{2}{3} \quad (D.5)$$

To estimate (B.15), let us propose that the relative amounts $X_j(t_0)$ (2.4) invested into securities $j=1,2$ of the market portfolio take the values:

$$X_1(t_0) = \frac{1}{3} \ ; \quad X_2(t_0) = \frac{2}{3} \quad (D.6)$$

From (D.6), obtain the estimate of Markovitz variance $\Theta_M(t,t_0)$ (D.4) of the market portfolio:

$$\Theta_M(t, t_0) = \frac{1}{9} \theta_{11}(t, t_0) + \frac{4}{9} \theta_{22}(t, t_0) \quad (D.7)$$

From (D.5-D.7) and (B.15), obtain the estimate of Markovitz variance $\Theta_{Mm}(t,t_0)$ (D.3) of the trades with all market securities:

$$\Theta_{Mm}(t, t_0) = \frac{1}{36} \theta_{11}(t, t_0) + \frac{16}{9} \theta_{22}(t, t_0) \quad (D.8)$$

If one assumes that the variances $\theta_{11}(t,t_0)$ and $\theta_{22}(t,t_0)$ of securities *1* and *2* are almost the same:

$$\theta_{11}(t, t_0) \sim \theta_{22}(t, t_0) \quad (D.9)$$

then one may compare $\Theta_M(t,t_0)$ (D.7) and $\Theta_{Mm}(t,t_0)$ (D.8):

$$\Theta_M(t,t_0) \sim \frac{20}{36} \theta_{11}(t,t_0) \quad ; \quad \Theta_{Mm}(t,t_0) \sim \frac{65}{36} \theta_{11}(t,t_0) \quad \rightarrow \quad \Theta_{Mm} \sim 3 \cdot \Theta_M(t,t_0) \qquad (D.10)$$

Our toy model reveals that if all volumes $U_j(t_i)$ of consecutive trades with securities are constant, then Markowitz variance $\Theta_M(t,t_0)$ (D.7) of the market portfolio may be three times less (D.10) than Markovitz variance $\Theta_{Mm}(t,t_0)$ (D.8) of all trades in the market. In other words, we demonstrate that the particular values (D.5; D.6) may cause the risks of the trades in the market to be three times more than the assessment of the risks of the market portfolio via Markowitz variance.

It is obvious that the different values of the relative numbers $x_j(t_0)$ (2.3) and $x_j(t)$ (4.4) and the different relative amounts $X_j(t_0)$ (2.4) invested into securities $j=1,2$ could give opposite result.

$$x_1(t_0) = \frac{W_1(t_0)}{W_\Sigma(t_0)} = \frac{1}{5} \; ; \; x_2(t_0) = \frac{4}{5} \; ; \; x_1(t) = \frac{U_1}{U_m} = \frac{3}{5} \; ; \; x_2(t) = \frac{2}{5} \qquad (D.11)$$

$$X_1(t_0) = 0{,}1 \; ; \quad X_2(t_0) = 0{,}9 \qquad (D.12)$$

$$\Theta_M(t,t_0) = 0{,}01\, \theta_{11}(t,t_0) + 0{,}81 \theta_{22}(t,t_0) \; ; \; \Theta_{Mm}(t,t_0) = 0{,}09 \cdot \theta_{11}(t,t_0) + \frac{0{,}81}{4} \theta_{22}(t,t_0)$$

If we also assume (D.9), then one obtains:

$$\Theta_M(t,t_0) \sim 0{,}82\, \theta_{11}(t,t_0) \; ; \; \Theta_{Mm}(t,t_0) = 0{,}29 \cdot \theta_{11}(t,t_0) \qquad (D.13)$$

Thus, for the case (D.11; D.12), obtain the relations that are almost opposite to (D.10):

$$\Theta_M(t,t_0) \sim 2{,}8 \cdot \Theta_{Mm}(t,t_0) \qquad (D.14)$$

The simplest toy model with constant volumes $U_j(t_i)$ of consecutive trades with securities of consecutive trades with securities reveals that the relations between Markowitz variances of the market portfolio and the trades with all securities in the market may vary a lot.

Now we briefly consider the toy model where the volumes $U_m(t_i)$ of consecutive trades with all securities in the market are constant, but the volumes $W(t_i)$ of trades with the market portfolio are random.

Let us assume that at time $t_i$ the volumes $U_j(t_i)$ of trades with securities $j=1,2$ obey the relations:

$$U_1(t_i) = U_1 + \delta(t_i) \geq 0 \; ; \; U_2(t_i) = U_2 - \delta(t_i) \geq 0 \; ; \; U_1 \leq U_2 \qquad (D.15)$$

The volumes $U_1$ and $U_2$ are constant and $\delta(t_i)$ is a random value with a zero mean $\delta(t)$ (D.16):

$$\delta(t) = \frac{1}{N}\sum_{i=1}^{N} \delta(t_i) = 0 \quad ; \quad \alpha^2 = \frac{1}{N}\sum_{i=1}^{N} \delta^2(t_i) \; ; \; \alpha^2 < U_1^2 \qquad (D.16)$$

Due to (D.15), the volumes $U_m(t_i)$ (4.1) of trades are constant:

$$U_m(t_i) = U_1 + \delta(t_i) + U_2 - \delta(t_i) = U_m = U_1 + U_2 \qquad (D.17)$$

Meanwhile, the volumes $W(t_i)$ (3.4; D.1; D.2) of trades with the market portfolio are random:

$$W(t_i) = \sum_{j=1}^{2} \frac{W_j(t_0)}{U_{\Sigma j}(t)} U_j(t_i) = W + w\delta(t_i) \; ; \; W = \frac{1}{N}[W_1(t_0) + W_2(t_0)] \; ; \; w = \frac{W_1(t_0)}{U_{\Sigma 1}(t)} - \frac{W_2(t_0)}{U_{\Sigma 2}(t)} \quad (D.18)$$

If $w\neq 0$ (D.18), then:

$$W(t;1) = W \quad ; \quad \delta W(t_i) - W(t;1) = w\delta(t_i) \tag{D.19}$$

The coefficient of variation $\chi^2$ (5.12) of the volumes $W(t_i)$ takes the form:

$$cov\{W(t_i), W(t_i)\} = w^2\alpha^2 \quad ; \quad \chi^2 = \frac{cov\{W(t_i), W(t_i)\}}{W^2} = \frac{w^2\alpha^2}{W^2} \tag{D.20}$$

The use of the coefficient $\gamma(t_i)$ (C.1; C.9) for the case (D.15-D.17) gives the same result:

$$W(t_i) = \gamma(t_i)U_m \quad ; \quad \gamma(t) = \frac{W}{U_m} \quad ; \quad cov\{\gamma(t_i), \gamma(t_i)\} = \frac{w^2\alpha^2}{U_m^2} \quad ; \quad \chi_\gamma^2 = \frac{w^2\alpha^2}{W^2} = \chi^2 \tag{D.21}$$

The Taylor series of the variances $\Theta(t,t_0)$ (7.4; D.22) of the market portfolio depend on the coefficient of variation $\chi^2$ (D.20) of the volumes $W(t_i)$ (3.4; D.18) of trades with the market portfolio.

$$\Theta(t,t_0) = (\psi_0^2 + \chi^2) \cdot R^2(t,t_0) = \Theta_M(t,t_0) + \chi^2 \cdot R^2(t,t_0) \tag{D.22}$$

The growth of the coefficient of variation $\chi^2$ (D.20) may increase the variance $\Theta(t,t_0)$ (D.22) of the market portfolio to compare with its Markowitz variance $\Theta_M(t,t_0)$. Thus, the impact of the coefficient of variation $\chi^2$ (D.20) of the volumes $W(t_i)$ (3.4; D.18) of trades with the market portfolio may increase the variance $\Theta(t,t_0)$ (D.22) and the risks of the market portfolio and its distinctions from the variance $\Theta_{Mm}(t,t_0)$ (D.13) of all trades in the market.

The investors and the portfolio managers should keep in mind the possibility of such a high distinction between the variances and the risks of the market portfolio and of trades with all securities in the market. The measurements and the forecasting of their variances may help achieve more reliability of the investments in market portfolios. That requires further studies.